\documentclass[10pt,journal,compsoc]{IEEEtran}
\ifCLASSOPTIONcompsoc
  \usepackage[nocompress]{cite}
\else
  \usepackage{cite}
\fi
\ifCLASSINFOpdf
  \usepackage[pdftex]{graphicx}
  \graphicspath{figures/}
  \DeclareGraphicsExtensions{.pdf,.jpg,.png}
\else
  \usepackage[dvips]{graphicx}
  \graphicspath{./figures/eps/}}
  \DeclareGraphicsExtensions{.eps}
\fi
\usepackage{amsmath,amssymb}
\usepackage{algorithmic}
\ifCLASSOPTIONcompsoc
  \usepackage[caption=false,font=footnotesize,labelfont=sf,textfont=sf]{subfig}
\else
  \usepackage[caption=false,font=footnotesize]{subfig}
\fi
\usepackage{wrapfig}

\usepackage{xcolor}
\usepackage{pict2e}
\usepackage{tikz}
\usepackage{algorithm}

\newcommand{\squishlist}{
   \begin{list}{$\bullet$}
    { \setlength{\itemsep}{0pt}      \setlength{\parsep}{3pt}
      \setlength{\topsep}{3pt}       \setlength{\partopsep}{0pt}
      \setlength{\leftmargin}{1.5em} \setlength{\labelwidth}{1em}
      \setlength{\labelsep}{0.5em} } }

\newcommand{\squishlisttwo}{
   \begin{list}{$\bullet$}
    { \setlength{\itemsep}{0pt}    \setlength{\parsep}{0pt}
      \setlength{\topsep}{0pt}     \setlength{\partopsep}{0pt}
      \setlength{\leftmargin}{2em} \setlength{\labelwidth}{1.5em}
      \setlength{\labelsep}{0.5em} } }

\newcommand{\squishend}{
    \end{list}  }

\newcommand{\myparagraph}[1]{\noindent\textbf{#1}\,\,}

\begin{document}
\title{GLoG: Laplacian of Gaussian for Spatial Pattern Detection in Spatio-Temporal Data}

\author{Luis~Gustavo~Nonato,~\IEEEmembership{Member,~IEEE,}
Fabiano~Petronetto,
and~Claudio~Silva,~\IEEEmembership{Member,~IEEE}% <-this % stops an unwanted space
\IEEEcompsocitemizethanks{\IEEEcompsocthanksitem Gustavo Nonato is with University of S\~ao Paulo - Brazil.\protect\\
E-mail: gnonato@icmc.usp.br.\protect\\
\IEEEcompsocthanksitem Fabiano Petroneto is with Federal University of Esp\'{\i}rito Santo - Brazil.\protect\\
E-mail: fabiano.carmo@ufes.br.\protect\\
\IEEEcompsocthanksitem Claudio Silva is with New York University.\protect\\
E-mail: csilva@nyu.edu.}%
%\thanks{Manuscript received April 19, 2005; revised August 26, 2015.}%
}

\markboth{Journal of \LaTeX\ Class Files,~Vol.~14, No.~8, August~2015}%
{Shell \MakeLowercase{\textit{et al.}}: Bare Demo of IEEEtran.cls for Computer Society Journals}



\IEEEtitleabstractindextext{%
\begin{abstract}
Boundary detection has long been a fundamental tool for image processing and computer vision, supporting the analysis of static and time-varying data. In this work, we built upon the theory of Graph Signal Processing to propose a novel boundary detection filter in the context of graphs, having as main application scenario the visual analysis of spatio-temporal data. More specifically, we propose the equivalent for graphs of the so-called Laplacian of Gaussian edge detection filter, which is widely used
in image processing. The proposed filter is able to reveal interesting spatial patterns while still enabling the definition of entropy of time slices. The entropy reveals the degree of randomness of a time slice, helping users to identify expected and unexpected phenomena over time. The effectiveness of our approach appears in applications involving synthetic and real data sets, which show that the proposed methodology is able to uncover interesting spatial and temporal
phenomena. The provided examples and case studies make clear the usefulness of our approach as a mechanism to support visual analytic tasks involving spatio-temporal data.
\end{abstract}

\begin{IEEEkeywords}
Data Filtering, Data Transformation, Feature Detection.
\end{IEEEkeywords}}

\maketitle

\IEEEdisplaynontitleabstractindextext

% \fi
\IEEEpeerreviewmaketitle

\IEEEraisesectionheading{\section{Introduction}\label{sec:introduction}}

%\IEEEPARstart{T}{his} demo file is intended to serve as a ``starter file''
%for IEEE Computer Society journal papers produced under \LaTeX\ using
%IEEEtran.cls version 1.8b and later.
%I wish you the best of success.

Feature extraction and transformation comprise fundamental steps
in the visualization and visual analytic pipeline. 
In the particular case of spatio-temporal data, such preprocessing steps
are of critical importance, as many visual analytic tools
rely on features to enable meaningful visualizations of patterns and trends hidden 
on the data~\cite{andrienko:tvcg:2017,liu2018tpflow,miranda:tvcg:2017}.
Over the years, a number of techniques devoted to extract and processing
features from spatio-temporal data have been proposed, 
ranging from simple aggregation schemes~\cite{andrienko:book:2006} to sophisticated topological~\cite{doraiswamy:tvcg:2014} and
tensor decomposition methods~\cite{cao2018voila}. 
Those techniques are designed to capture specific properties or phenomena present in the data.
For instance, topological methods look for locations where data assume extreme
values while techniques based on tensor decomposition aim to decompose the data so
as to identify regions and time slices with common properties.  

Identifying locations where data changes abruptly has long been the goal
of many feature extraction methods.  
Regions of sharp variation of the data typically correspond to locations where data transitions
from one ''state'' to another, thus being of great relevance for analytical purposes. 
Feature extraction methods able to detect local changes in a signal have
been extensively used in fields such as image processing and computer vision~\cite{shapiro2001computer}, 
being called \emph{edge detection} methods.
In fact, edge detection has been the building block of methods designed for recognizing and segmenting objects, enhancing details, and performing feature preserving denoising in images and videos. 
Despite the importance, few has been done towards developing techniques similar to edge detection to assist visual analytic tasks, 
mainly in applications involving data defined in unstructured domains.
A reason for this gap is that most edge detection techniques developed in the context of image processing rely on mathematical tools such as derivatives and convolution, which can hardly be defined on non-regular grid-like domains.

This work brings an alternative to the issue pointed above, proposing an operator able to identify abrupt changes 
in a signal defined on the nodes of a graph. 
More specifically, we propose a novel filter, called \emph{GLoG},
which is the counterpart for graphs of the so-called Laplacian of Gaussian edge 
detection method (also known as Marr-Hildreth operator) widely used in image processing~\cite{maini2009study}.
Our approach relies on the theory of Graph Signal Processing (GSP)~\cite{shuman2013emerging}, which
provides a solid mathematical framework for adapting and extending well known tools from the classical signal processing field 
to the more general context of graphs~\cite{dal2017wavelet,valdivia:vast:2015}.
The proposed GLoG filter can identify spatial locations of abrupt changes in a signal, uncovering regions (boundaries) 
where the signal changes its properties.
Moreover, we build upon the GLoG filter to highlight time intervals where ``expected'' and ``unexpected'' boundaries take place. 
In other words, we rely on the GLoG filter to 
define the concept of \emph{entropy} of time slices, from which we derive a \emph{temporal entropy diagram}. 
The latter allows the visual identification of time instances  
where observed boundaries are likely to happen (lower entropy time instants) as well as 
moments where observed boundaries are less expected (higher entropy time instants).
The resulting analysis makes easier the visual identification of expected and unexpected patterns over time.
Nevertheless, the Gestalt law of proximity \cite{ware2012information} states that groups of points spatially close to each other
are pre-attentively perceived as a common set of abstract features. The boundaries extracted from a signal fit naturally
in this concept, turning out a valuable visual resource to reveal spatial patterns and trends. 

We show the effectiveness of our approach in synthetic and real data sets containing
information about taxi trips in Manhattan, NYC, and geo-referenced criminal activities in the city of S\~ao Paulo, Brazil.
The provided examples and case studies make clear the usefulness of the proposed boundary detection and 
entropy diagram as basic tools to support the visual analysis of spatio-temporal data.

In summary, the main contributions of this work are:
\squishlist
\vspace{-.09cm}
\item \emph{GLoG} (Graph Laplacian of Gaussian), a boundary detection filter
    that is the graph counterpart of the so-called Laplacian of Gaussian filter typically
    used in image processing. As far as we know, this is the first time a boundary detection filter 
    is proposed in the context of Graph Signal Processing;
\vspace{-.09cm}
\item  the concept of entropy of time slices and associated entropy diagram, which enables the visual identification 
    of expected and unexpected spatial phenomena over time;
\vspace{-.09cm}
\item two case studies that attest the usefulness and potential of the
    proposed methodology to support visualization assisted analysis of spatio-temporal data.
    These case studies shows the potential of GLoG operator as a feature extraction method 
    in the context of spatio-temporal data.
\squishend
\vspace{-.09cm}
We emphasize that the proposed GLoG filter should not be seen as a competitor of other
feature extraction techniques. In fact, the GLoG aims to detect a specific property (abrupt change) of a signal,
which is not captured by existing methods. Therefore, GLoG can be combined with other techniques 
to produce features that captures a wide variety of traits present in spatio-temporal data.

%-------------------------------------------------------------------------
\vspace{-.35cm}
\section{Related Work}
\vspace{-.05cm}
The literature about visualization assisted spatio-temporal data analysis is extensive, ranging from  
georeferenced time-varying data analysis~\cite{andrienko:tvcg:2017,candela2018radian} to dynamic networks~\cite{dal2018wavelet,van2016reducing}.
The different methodologies have been reviewed and organized
in a number of surveys~\cite{an2015,andrienko2017visual,bach:eurovis:2014,chen2015survey} 
and books~\cite{andrienko:book:2006,cressie:book:2015}. 
To better contextualize our contribution, we focus on techniques
that rely on feature extraction and data transformation mechanisms to leverage the visualization of
spatial and temporal phenomena present in the data.
We group spatio-temporal feature extraction and/or data transformation techniques in three categories: 
temporal, spatial, and spatio-temporal. 

\smallskip
\noindent\textbf{Temporal} methods aim to uncover patterns by transforming the temporal 
component of the data associated in each spatial locations. 
In visual analysis, transformations such as temporal aggregation~\cite{Andrienko:sigspatial:2013b,chen:tvcg:2016} and
moving averages~\cite{itoh:bigdata:2014} figure among the most common temporal approaches.
There are also alternatives that rely on prediction mechanisms to detect unexpected temporal events, 
which are highlighted during visualization~\cite{xia:www:2014}.
Classical signal processing methods have also been employed to extract features from time-series associated 
to specific spatial locations so as to make the visual identification of similar temporal patterns an easier 
task~\cite{zhang:kdd:2015}.
When regions made up of several locations have to be handled, 
spatial data aggregation is firstly employed to combine multiple time-series into a single one, which is then 
processed to identify patterns and trends~\cite{andrienko2010general}.
Aggregation tends to attenuate high-frequency events, hampering the identification of patterns associated to abrupt variations 
of the signal in particular locations. In fact, aggregation can be seen as low-pass filter applied 
prior to a feature extraction mechanism. The approach proposed in this work naturally combines
low pass filter and feature extraction in a single operator.

\smallskip
\noindent\textbf{Spatial} transformation schemes aim to uncover spatial patterns present in the data. 
More specifically, given a time slice (which can correspond to aggregated data from 
a time interval), spatial methods typically split the spatial domain in subregions, 
extracting and/or emphasizing features in each subregion. 
Subregions can be defined based on density information~\cite{Scheepens:tvcg:2011} 
or by grouping spatial locations with similar content~\cite{andrienko:tvcg:2017,thakur:iv:2010,wu:tvcg:2016}. 
A large number of methods have been proposed to extract features from spatial locations to support visualization tasks.
Some examples are the interchangeable matrices~\cite{zeng:cgf:2013}, which
uncover co-occurring spatial events, and topological approaches~\cite{doraiswamy:tvcg:2014,miranda:tvcg:2017}, 
which identify spatial locations that bear ``peculiar'' behavior. 
Topic modeling has been exploited as a mechanism to characterize 
spatial locations~\cite{chu:pacificvis:2014}.
Graph signal processing tools such as windowed graph Fourier transform have been employed to
extract features from spatial locations~\cite{shuman:acha:2016}.
Most of the spatial methods described above are designed to extract and identify
patterns in fixed time slice, resorting to visualization as a main analytical resource to understand  patterns
and their dynamics over time. The boundary detection approach proposed in the current work 
can be seen as a spatial method, although the derived entropy diagram is a temporal visual tool.

\begin{figure*}
\centering
\includegraphics[width=0.99\linewidth]{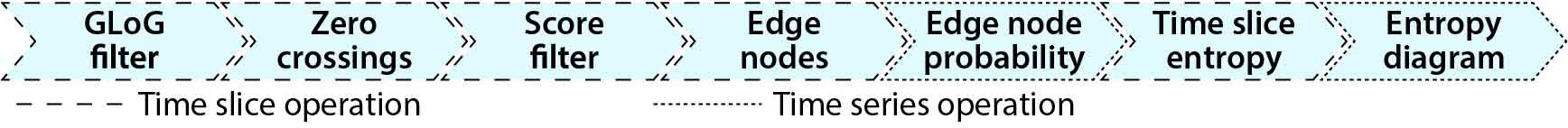}
    \caption{Method pipeline: our methodology apply GLoG filter at each graph signal time slice to compute edge node configurations and define entropy diagram.}\vspace{-0.5cm}
    \label{fig:pipe}
\end{figure*}
\smallskip
\noindent\textbf{Spatio-Temporal} transformation techniques operate on temporal and spatial 
dimensions so as to extract meaningful information that is emphasized during visualization. 
A typical alternative is to compute separate temporal and spatial features, resorting to 
visual analytic tools to unify the spatial and temporal analysis. This is the methodology 
implemented in the MobilityGraphs technique~\cite{vonLandesberger:tvcg:2016},
where spatial or temporal signatures are used to cluster trajectories in order to reduce data complexity
and visual clutter. 
Another common use of independent spatial and temporal transformations 
is the use of spatial signatures to identify regions of interest and
temporal features to analyze the behavior of such regions over time~\cite{maciejewski:tvcg:2010}.
Some techniques dynamically extract temporal and spatial features according to 
user specified queries, enabling complex analysis of time-stamped trajectories~\cite{giannotti:kdd:2007,lorenzo:tvcg:2016}, 
frequent trips~\cite{yu:vast:2015}, and spatial/temporal 
traffic patterns~\cite{ferreira:tvcg:2013,guo:pvis:2011}.

More elaborate techniques operate on spatial and temporal information simultaneously, 
allowing for identifying spatial, temporal, and spatio-temporal patterns. 
Feature vectors made up of temporal and spatial attributes is an alternative 
that have been applied in several scenarios~\cite{zhang:tvcg:2014}.
Optimization procedures designed to detect periodical patterns in spatial events 
have also been employed to visually identify tends in epidemiological data~\cite{matsubara:kdd:2014}.
Spatio-temporal wavelet-based signatures~\cite{valdivia:vast:2015}
have turned out effective to simultaneously characterize spatial locations and their temporal behavior.
More recently, transformations based on tensor decomposition are being used 
as a resource to assist the visualization of spatio-temporal data~\cite{cao2018voila,liu2018tpflow}.

The proposed feature extraction and transformation techniques bear a 
number of properties not simultaneously present in any existing technique. Similarly to topological methods 
for spatio-temporal analysis~\cite{doraiswamy:tvcg:2014,miranda:tvcg:2017},
the proposed approach is scale invariant, thus producing identical spatial ``signatures'' for signals that only differ
by a scaling factor. However, the capability of generating spatial signatures on regions of large variation of the 
signal is a differential of our approach when compared to topological schemes. Topological approaches 
are able to point out the location where a signal reaches its extremal values, but it demands very
sophisticated methods (like Morse-Smale complexes computation) to account for where 
the underlying phenomenon is transitioning, a trait naturally captured by our approach. 
Moreover, the proposed methodology allows for transforming spatial 
patterns into a single scalar value (entropy) that indicates the degree of randomness of each 
time slice. Entropy values are much easier to interpret than, for example, patterns derived from tensor decomposition,
which typically demand sophisticated visualization resources to become meaningful~\cite{cao2018voila,liu2018tpflow}. 

%------------------------------------------------------------------------
\vspace{-.35cm}
\section{Edge Node Configuration and Entropy Diagram to Support Visual Analytic Tasks}
\vspace{-.05cm}
The proposed spatio-temporal analytic tool comprises two main steps, i) the detection of boundary nodes in each time slice  
and ii) the computation of entropy for the time slices, as illustrated in Figure~\ref{fig:pipe}.
The former relies on the GLoG operator described in Section~\ref{sec:glog} while the computation of 
entropy derives from the concept of edge node probability discussed in Section~\ref{sec:tse}.
Before presenting both concepts, though, we describe 
the mathematical foundations of Graph Fourier Transform (GFT)
and Graph Filtering, which are the basis of our approach.

%-------------------------------------------------------------------------
\vspace{-.25cm}
\subsection{Graph Signal Processing: Basic Concepts} 
\vspace{-.15cm}
\label{sec:gsp}
This section presents basic concepts of 
Graph Signal Processing from which our approach derives.
A more detailed discussion about basic concepts of Graph Signal Processing
can be found in the work by Shuman et al.~\cite{shuman2013emerging}.

\smallskip
\myparagraph{Graph Fourier Transform.}
We denote by $\mathcal{G}=(V,E,w)$ a \emph{graph} made up of a node
set $V=\{\tau_1,\tau_2,\ldots,\tau_n\}$, an edge set
$E=\{(\tau_i,\tau_j),\, \tau_i,\tau_j\in V,\,i\not=j\}$, and a
weight function $w: E \rightarrow \mathbb{R}$ that associates a non-negative 
scalar to each edge in $\mathcal{G}$. 
In our context, $\mathcal{G}$ is assumed to be \emph{connected}, that is, for every pair of nodes there
is a sequence of adjacent edges connecting those nodes.

The \emph{weighted adjacency matrix} of $\mathcal{G}$, denoted by
$A=(a_{ij})$, is the matrix satisfying $a_{ij}=w(\tau_i,\tau_j)$ if $(\tau_i,\tau_j)\in
E$ and $a_{ij}=0$ otherwise. This matrix is used to define the
(non-normalized) \emph{graph Laplacian}, which is given by $L=D-A$, where
$D=\text{diag}(d_1,d_2,\ldots,d_n)$ is a diagonal matrix with entries
$d_i=\sum_j a_{ij}$ and $n$ is the number of nodes in $V$.  The graph Laplacian is a real, symmetric, and
semi-positive definite matrix, which ensures a complete set of
orthonormal real eigenvectors $u_\ell$, with corresponding non-negative
real eigenvalues $\lambda_\ell$, $\ell=1,2,\ldots,n$. 
Moreover, zero is always an eigenvalue of $L$ whose corresponding 
eigenvector is a constant vector.

The eigenvalues and eigenvectors of the graph Laplacian play a similar role as 
frequencies and basis functions in the classical Fourier theory. More specifically,
eigenvalues closer to zero correspond to low frequencies
while large eigenvalues correspond to high frequencies. 
Moreover, eigenvectors associated to 
small eigenvalues tend to have a less oscillatory behavior than eigenvectors associated 
to large eigenvalues. A more detailed discussion about the relation between the spectrum of 
Laplacian matrices and Fourier theory can be found in the work by Shuman et al.~\cite{shuman2016vertex}
and Dal Col et al.~\cite{dal2017wavelet}.

A signal defined on the nodes of $\mathcal{G}$ is a
function $f:V\rightarrow\mathbb{R}$ that associates a scalar $f(\tau_i)$ to each node $\tau_i\in V$.  
Denoting the eigenvalues and corresponding eigenvectors of the Laplacian matrix of $\mathcal{G}$ 
by $\lambda_\ell$ and $u_\ell$, $\ell=1,\ldots,n$ respectively,
and assuming the eigenvalues are sorted in non-decreasing order,
$0=\lambda_1<\lambda_2\leq\ldots\leq\lambda_n$ (the first strict inequality is due to the assumption that $\mathcal{G}$ is connected), 
the \emph{Graph Fourier Transform} (GFT) of a signal $f$, denoted by
$\widehat{f}:\Lambda\rightarrow\mathbb{R}$, where $\Lambda$ is the spectral domain (set of eigenvalues), is defined as:
\begin{equation}
  \widehat{f}(\lambda_\ell) = \langle u_\ell,f \rangle = \displaystyle\sum_{j=1}^{n} u_\ell(\tau_j)f(\tau_j),
  \label{eq:gft}
\end{equation}

Given the GFT $\widehat{f}$, the original signal $f$ can be recovered via the 
\textit{inverse Graph Fourier Transform} (iGFT), which is defined as:
\begin{equation}
    f = iGFT(\widehat{f}) = \displaystyle\sum_{\ell=1}^{n} \widehat{f}(\lambda_\ell)u_\ell
\label{eq:igft}
\end{equation}
If we denote by $U$ the (orthogonal) matrix with columns given by the eigenvectors $u_\ell$,
the GFT and iGFT can be obtained via matrix multiplication as follows:
\begin{eqnarray}
\fbox{\mbox{GFT}}\,\,\, & \qquad & \,\,\,\fbox{\mbox{iGFT}}\nonumber\\
\widehat{f} = U^{\!\top}f & \qquad & f = U\widehat{f} 
\label{eq:mgf}
\end{eqnarray}

\smallskip
\myparagraph{Spectral Filtering.}
A \emph{graph spectral filter} 
$\widehat{h}:\Lambda\rightarrow\mathbb{R}$ is a function defined in the
spectral domain that associates a scalar value $\widehat{h}(\lambda_\ell)$
to each eigenvalue $\lambda_\ell\in\Lambda$. 
The GFT $\widehat{f}:\Lambda\rightarrow\mathbb{R}$ can be seen as a particular
instance of a graph spectral filter.  

A \emph{graph spectral filtering} of a signal $f$ is defined as:
\begin{equation}
\widehat{f}_h = \widehat{f}\cdot\widehat{h}
\label{eq:gsf}
\end{equation}
where $\widehat{f}$ is the GFT of $f$, $\widehat{h}$ is a graph spectral filter and $\cdot$ is the element-wise multiplication.
Using the matrix notation defined in Eq.~\eqref{eq:mgf} and some algebraic manipulation
one can obtain the filtered version $\widetilde{f}$ of $f$ in the graph domain by computing:
\begin{equation}
\widetilde{f}=UHU^{\!\top}f
\label{eq:gf}
\end{equation}
where $H$ is a diagonal matrix with entries $\widehat{h}(\lambda_1),\ldots,\widehat{h}(\lambda_n)$.

\begin{figure}[!t]
\centering
    \includegraphics[width=0.9\linewidth]{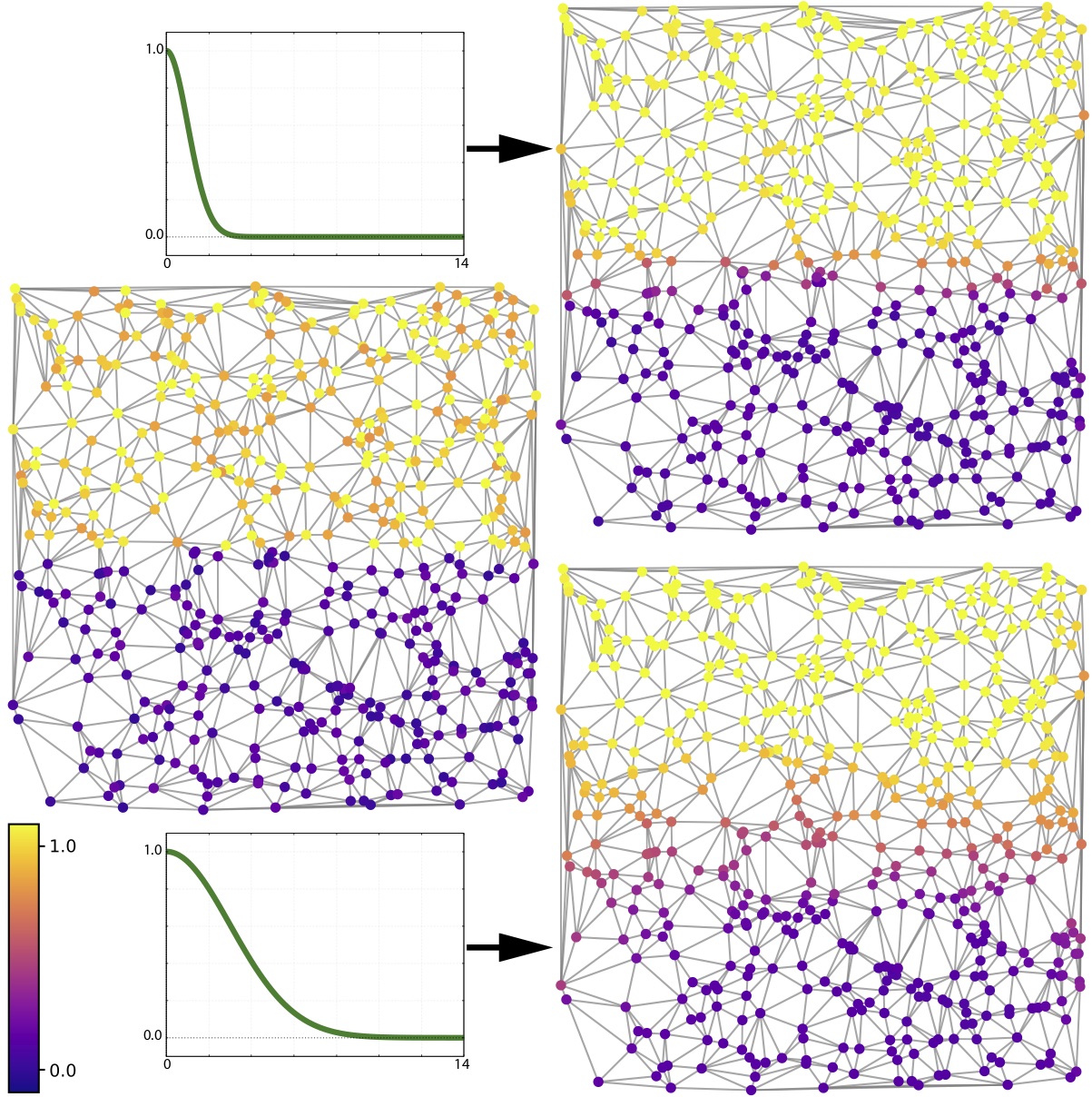}
    \caption{Noisy step function (left) $f(\tau_i) = 1 + \epsilon$ in the yellowish nodes  
    and $f(\tau_i) = \epsilon$ in the purplish, where $\epsilon$ is a random noise. 
    Gaussian smoothing with $\sigma=1$ (top right) and $\sigma=3$ (bottom right).
    The two plots on the left shows the shape of the Gaussian filter related.} \vspace{-.5cm}
    \label{fig:filters}
\end{figure}

The design a proper filter $\widehat{h}$ is application dependent. A particularly useful filter in our context is the  
Gaussian filter given by:
\begin{equation}
    \widehat{h}_g(\lambda)  =  \frac{1}{\sigma\sqrt{2\pi}}\exp \left(\frac{-\lambda^2}{2\sigma^2}\right) 
\end{equation}
where $\lambda$ is the independent variable and $\sigma$ is a parameter. The Gaussian filter is a low-pass filter
that preserves low frequencies while attenuating the higher ones.
Figure~\ref{fig:filters} illustrates the effect of applying $\widehat{h}_g$ to a noisy step function
with two different values of $\sigma$.

%-------------------------------------------------------------------------
\vspace{-.3cm}
\subsection{GLoG for Boundary Detection}
\label{sec:glog}

The boundaries (or edges) of a signal correspond to the locations of abrupt change in the signal. 
Many different approaches have been proposed to identify edges in the context of image processing and computer vision~\cite{maini2009study}.
Among the existing edge detection methods, the \emph{Laplacian of Gaussian} (LoG) figures among the most
important ones, mainly due to its solid mathematical foundation and optimality criteria~\cite{marr1980theory}.

\myparagraph{GLoG filter.} The classical LoG is a filter built based on two main principles: 1) boundaries take place on locations where 
the first derivative (or gradient) of a signal is maximum, or equivalently, the locations where
the Laplacian of the signal is zero, which is called the zero-crossings of the Laplacian;
2) zero-crossings may be caused by noise, so a smoothing filter must be applied to the signal before computing the Laplacian.
The chosen smoothing filter is the Gaussian filter, as it provides optimal localization in the space 
and frequency domains~\cite{marr1980theory}. 

In mathematical terms, the classical LoG filter can be defined as
\vspace{-.1cm}
\begin{equation}
    LoG(f) = \nabla^2(G*f) = \nabla^2 G*f
    \label{eq:log_def}
\end{equation}
where $\nabla^2$ is the Laplacian operator, $G$ is the Gaussian function, $f$ is the signal, 
and $*$ is the convolution operator. The right most expression is a consequence of the derivative rule for convolution.

There are two main issues for adapting Eq.\eqref{eq:log_def} in the context of graphs: how to define the convolution operator
and how to compute $\nabla^2 G$ on a graph structure.
The convolution operator is not well defined in graph domains, as it 
demands a shift mechanism that can not be directly defined on graphs. 
However, convolution becomes multiplication in the spectral domain, thus, 
with the help of Graph Signal Processing theory, the convolution between two functions $f$ and $g$ 
can be defined as~\cite{shuman2013emerging}:
\vspace{-.1cm}
\begin{equation}
    f*g = iGFT(\widehat{f}\cdot \widehat{g})
    \label{eq:conv}
\end{equation}
where $\cdot$ is the element-wise multiplication, $\widehat{f}$ and $\widehat{g}$ are the GFT of $f$ and $g$ 
(defined in Eq.~\eqref{eq:gft}), and $iGFT$ is the inverse GFT given in Eq.~\eqref{eq:igft}.
Using Equations~\eqref{eq:log_def} and \eqref{eq:conv}, we can define the \emph{Graph Laplacian of Gaussian} (GLoG) filter as:

\smallskip
\myparagraph{Definition (GLoG):}
\begin{equation}
    GLoG(f) = iGFT(\widehat{\nabla^2 G}\cdot\widehat{f}).
    \label{eq:log_gsp}
\end{equation}

\begin{figure}[!t]
\centering
\begin{tikzpicture}
\node[above right] (img) at (0,0) {\includegraphics[width=0.99\linewidth]{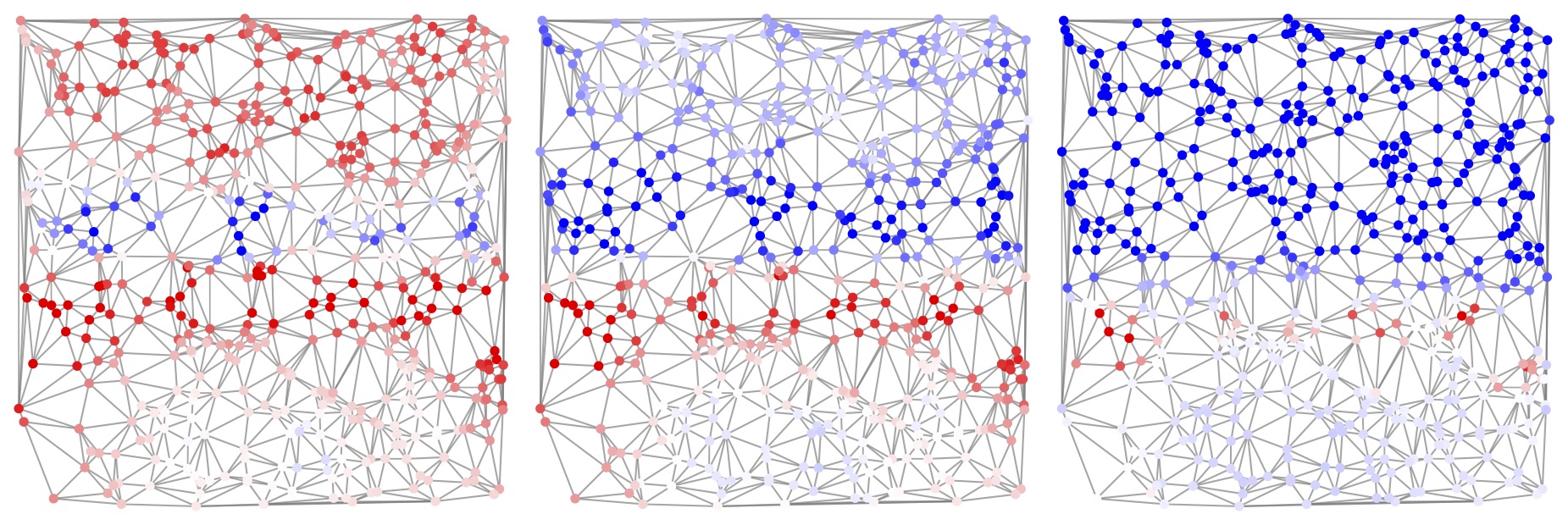}};
\node at (42pt,2pt) {$\sigma=1$};
\node at (124pt,2pt) {$\sigma=2$};
\node at (206pt,2pt) {$\sigma=3$};
\end{tikzpicture}
\vspace{-0.5cm}
    \caption{GLoG filter applied to the noisy step function depicted in 
    Figure~\ref{fig:filters} (left). Blue and red colors correspond to nodes where the GLoG filter is 
    negative and positive, respectively.\vspace{-0.5cm}}
    \label{fig:logs}
\end{figure}
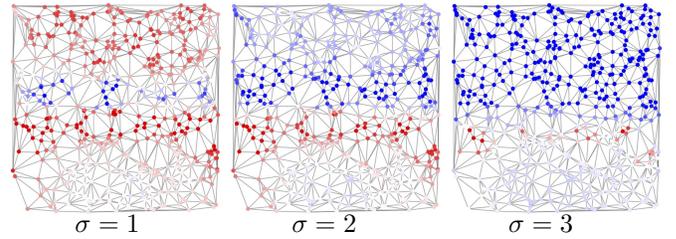
\begin{figure*}
  \centering
  \includegraphics[width=.975\linewidth]{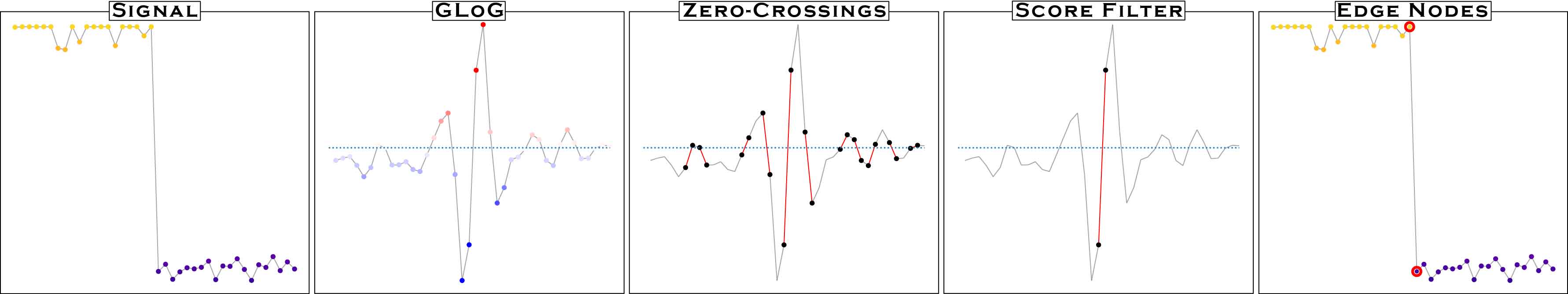}
  \caption{GLoG for boundary detection (Section~\ref{sec:glog}). From left to right, signal defined on a linear graph, GLoG of the signal, zero-crossing edges, the strongest edges, and the edge nodes obtained by algorithm (highlighted in red over graph signal).}
  \vspace{-.15cm}
  \label{fig:pipeline}
\end{figure*}

In order to make Eq.~\eqref{eq:log_gsp} of practical use, we must properly define $\widehat{\nabla^2 G}$.
We know that the classical 2D Fourier transform of $\widehat{\nabla^2 G}$ is given by $-4\pi^2\omega^2\exp(-\sigma^2\omega^2)$ ~\cite{marr1980theory}.
If we interpret the variable $\omega$, which is a complex variable in the classical Fourier transform theory, as a real
variable in the graph spectral domain, then $\widehat{\nabla^2 G}$ can be given by the following graph spectral filter:
\begin{equation}
    \widehat{\nabla^2 G}(\lambda) = -4\pi^2\lambda^2\exp(-\sigma^2\lambda^2).
    \label{eq:gft_log}
\end{equation}
where $\sigma$ is a user defined parameter and $\lambda$ is the independent variable in the graph spectral domain. 
Defining $\widehat{\nabla^2 G}$ as a graph spectral filter makes the definition and computation
of GLoG feasible, being one of the contributions of this work. 

Having defined $\widehat{\nabla^2 G}(\lambda)$,
the GLoG filter (Eq.~\eqref{eq:log_gsp}) becomes a band-pass filter in the spectral graph domain.
Moreover, in the particular case where $\mathcal{G}$ is a regular 2D grid, the GLoG 
matches the LoG filter as defined in classical signal processing.

Figure~\ref{fig:logs} shows the result of the GLoG filter when applied to the noisy step function 
depicted in Figure~\ref{fig:filters} (left) using three different values of $\sigma$. Blueish and reddish colors correspond to negative and
positive values of the GLoG respectively. Notice that the larger the value of $\sigma$ the smoother the
result of the GLoG filter is.

\begin{figure}[!t]
\centering
\begin{tikzpicture}
\node[above right] (img) at (0,0) {\includegraphics[width=.97500\linewidth]{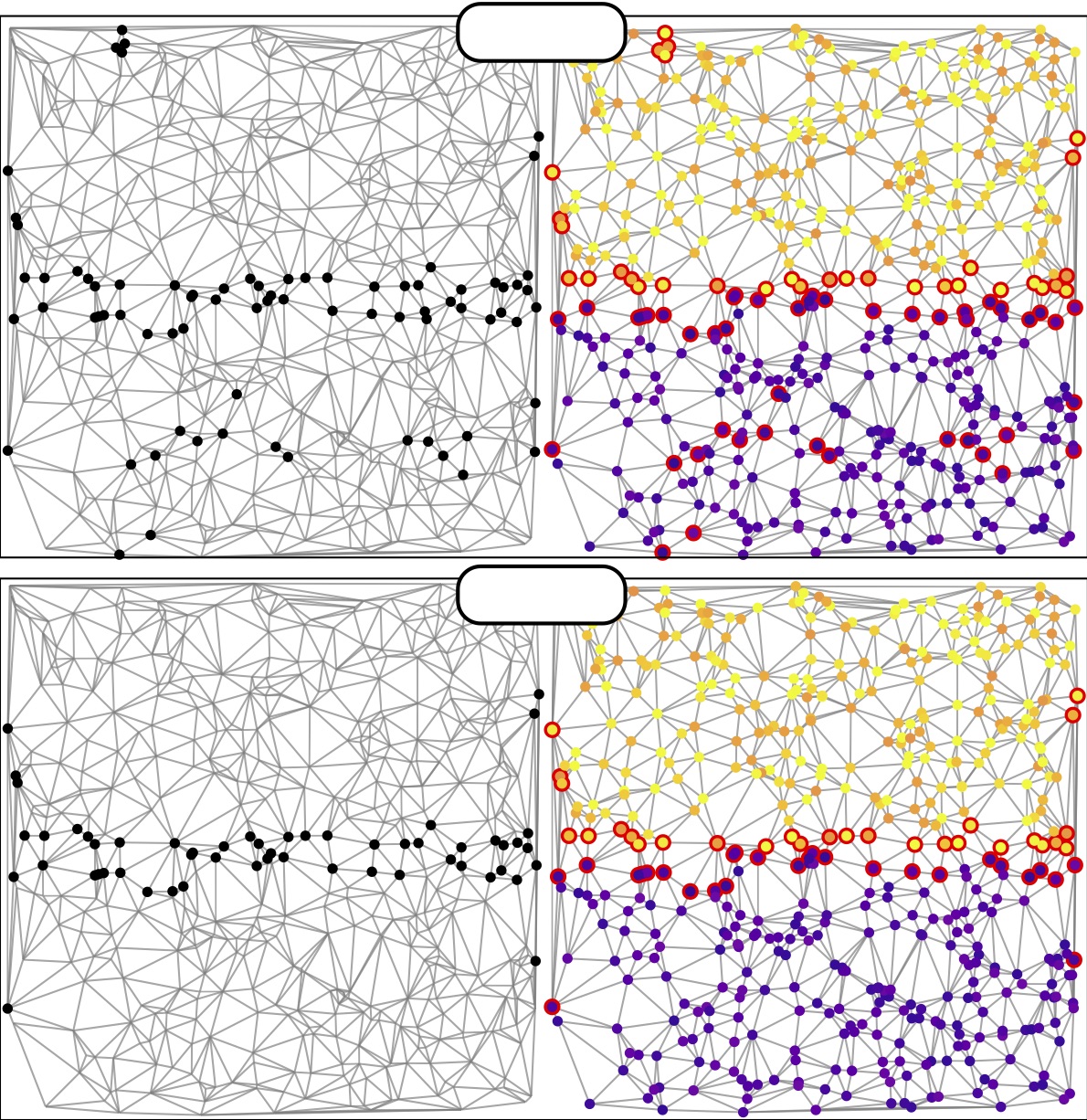}};
\node at (127pt,122pt) {\scalebox{0.9}{$\mu+2std$}};
\node at (127pt,250pt) {\scalebox{0.9}{$\mu+std$}};
\end{tikzpicture}
    \caption{Edge node configuration obtained by keep zero-crossing pairs 
    whose score is greater than the average $\mu$ plus (left) $1$
    and (right) $2$ times the standard deviation $std$ of the 
    scores in the $GLoG$ signals depicted in Figure~\ref{fig:logs} middle, respectively.}
    \vspace{-.15cm}
    \label{fig:step_e}
\end{figure}

\smallskip
\myparagraph{Extracting the Strongest Edge Nodes.} 
The result of applying a GLoG filter to a signal is another signal defined on the nodes of the graph, which we denote by $GLoG_f$ to make clear 
its dependence of the input signal $f$. A pair of adjacent nodes $(\tau_i,\tau_j)\in E$ 
is a \emph{zero-crossing pair} if $GLoG_f(\tau_i)GLoG_f(\tau_j)<0$. 
The nodes belonging to a zero-crossing pair are called \emph{edge nodes}.
We can associate the score $|GLoG_f(\tau_i)-GLoG_f(\tau_j)|$ to each 
zero-crossing pair, the larger the score the stronger the signal variation is in that pair, which we call \emph{strong edge nodes}
(or simply \emph{edge nodes}). 
``Weak'' pairs can be filtered out by considering only zero-crossing pairs whose score is among the largest ones. 
After computing the stronger edge nodes one can generate a binary signal $f_e$ where $f_e(\tau_i)=1$ if
$\tau_i$ is an edge node and $f_e(\tau_i)=0$ otherwise. The signal $f_e$ is called an \emph{edge node configuration} of $f$. 
Figure~\ref{fig:pipeline} illustrates all the steps involved in the {computation} of edge nodes of a signal $f$.

Figure~\ref{fig:step_e} shows edge node configurations for zero-crossing pairs whose score is greater
than the mean plus $1.0$ and $2.0$ times the standard deviation of the zero-crossing pair score distribution.
Such edges nodes have been computed from the GLoG signal for $\sigma=2$ depicted in Figure~\ref{fig:logs}.
Notice that, as expected, the higher the threshold the smaller the number of edge nodes.

\smallskip
\myparagraph{Computational Aspects.} The pseudocode for the GLoG filter computation can be stated as in Algorithm~\ref{alg:glog}. 
The method \texttt{get\_edge\_nodes} compute edge nodes by first identifying the edge node pairs,
that is, pairs of adjacent nodes where $g$ changes its sign. The edge node pairs whose score
is larger than a threshold (strong edge node pairs)  are returned by the method. 
In our implementation, the threshold is the third quartile of the edge node pair score distribution.
\begin{algorithm}
    \caption{GLoG($L$,$f$)}
        \begin{algorithmic}[1]
            \STATE Compute the eigenvector matrix $U$ and eigenvalues $\Sigma$ from the Laplacian $L$
            \STATE $\widehat{f} = U^{\!\top}f$ \hspace{3.4cm} \# GFT, Eq.~\eqref{eq:gft} 
            \STATE $\widehat{g} = \widehat{f}\cdot\widehat{\nabla^2 G}$ \hspace{2.9cm} \# $\widehat{\nabla^2 G}$ as in  Eq.~\eqref{eq:gft_log}
            \STATE $g = U\widehat{g}$ \hspace{3.64cm} \# iGFT, Eq.~\eqref{eq:igft}
            \STATE $e =$ \texttt{get\_edge\_nodes}($g$) \hspace{.9cm} \# find edge-nodes
            \RETURN $e$
        \end{algorithmic}
        \label{alg:glog}
\end{algorithm}
    
The spectral filtering can be computed via Chebyshev polynomial approximation~\cite{hammond2011wavelets},
which avoids the computation of the whole set of eigenvalues and eigenvectors of $L$,
thus making possible to handle large graphs in
reasonable computational times. If Chebyshev polynomial approximation is used, only
the largest eigenvalue has to be computed in step 1 of the algorithm and 
step 2 and 3 are merged in a single one (see \cite{hammond2011wavelets} for details).
Our implementation makes use of Chebyshev polynomial approximation.

%-------------------------------------------------------------------------
\subsection{Edge Node Probability and Time Slice Entropy}
\label{sec:tse}
In this section we will show how the boundary detection methodology described in the previous section 
can be used to assist spatio-temporal data visualization. 
Lets assume a time series is associated to each node of a graph $\mathcal{G}$, that is,
given a set of time slices $T=\{t_1,t_2,\ldots,t_m\}$,
there is a function $f:V\times T\rightarrow\mathbb{R}$ that associates a scalar value $f(\tau_i,t_j)$ to each
node $\tau_i$ in the time slice $t_j$.

\smallskip
\myparagraph{Edge Node Probability.} The GLoG filter \eqref{eq:log_gsp} can be applied to each time slice $t_j$ in order to identify the
edge nodes in $t_j$. Assuming that only the strongest edge nodes are kept in each time slice
(score larger than the third quartile of the score distribution), 
one can estimate the probability $p_e(\tau_i)$ of a node $\tau_i$ being an edge node as follows:
\begin{equation}
\label{eq:prob}
\begin{aligned}
    I(\tau_i,t_k) &= \left\{\begin{array}{ll}
            1 & \mbox{ if } \tau_i \mbox{ is an edge node in time slice } t_k \\
            0 & \mbox{ otherwise }
                \end{array}\right. \\
    p_e(\tau_i) &= \frac{1}{m}\sum_{k=1}^m I(\tau_i,t_k)
\end{aligned}
\end{equation}
where $m$ is the number of time slices.
The probability of a node $\tau_i$ not being an edge node is given by $1-p_e(\tau_i)$.
Notice that the probability $p_e(\tau_i)$ is simply the number of time slices where $\tau_i$
appears as an edge node divided by the total number of time slices.

\smallskip
\myparagraph{Time Slice Entropy.}
The following function
\begin{equation}
    p(I(\tau_i,t_k)) = \left\{\begin{array}{ll}
            p_e(\tau_i) & \mbox{ if } I(\tau_i,t_k) = 1 \\
            1-p_e(\tau_i) & \mbox{ if } I(\tau_i,t_k) = 0
                \end{array}\right. \\
    \label{eq:p}
\end{equation}
computes how probable the observed configuration of a node $\tau_i$ (in time slice $t_k$) is in terms of it being or not an 
edge node.
The \emph{entropy} of the edge node configuration in time slice $t_k$ is given by:

\smallskip
\noindent\textbf{Definition (Entropy):}
\begin{equation}
    E(t_k) = -\sum_{i=1}^n p(I(\tau_i,t_k))\log p(I(\tau_i,t_k))
    \label{eq:entropy}
\end{equation}
The entropy measures the degree of randomness of a time slice, the larger the entropy
the more unpredictable the time slice is in terms of its edge node configuration.
Time slice where the edge node configuration is of low probability presents larger entropy.
Therefore, by simply plotting the entropy over time, which we call \emph{entropy~diagram}, 
one can visualize time slices where the signal presents unexpected (high entropy) 
or predictable (low entropy) edge node configurations. The definitions of entropy and entropy diagram are
another contribution of this work.
The spatial distribution of edge nodes in each time slice is also an important feature
that can be used to understand a signal. 

%-------------------------------------------------------------------------
\subsection{Synthetic time-varying data}
\label{sec:synt}
\begin{figure}[!t]
\centering
    \includegraphics[width=0.975\linewidth]{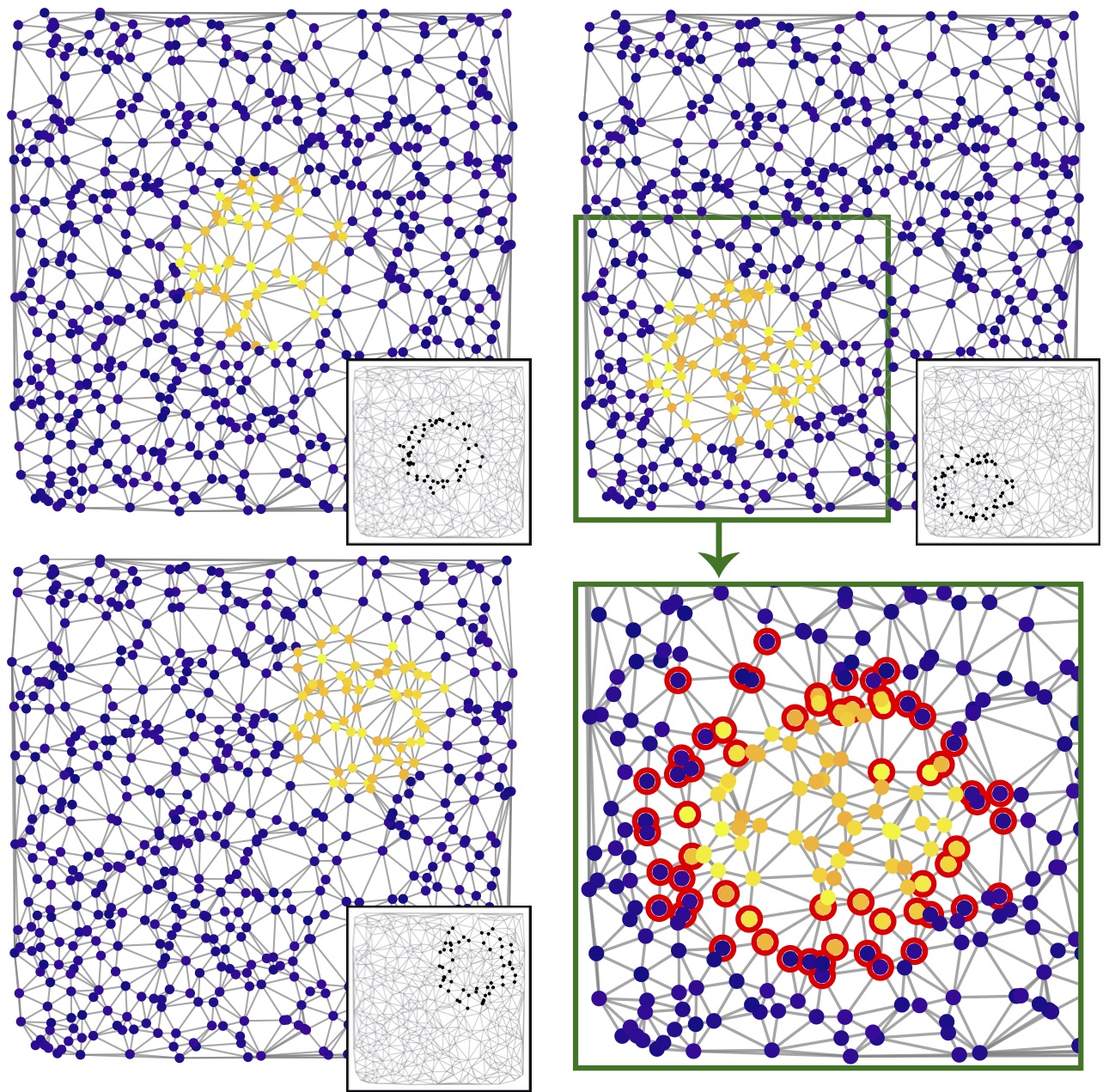}
    \caption{Three time slices of the synthetic signal and corresponding edge node configurations (botton right inset). 
    The bottom right image shows edge nodes (red circled nodes) jointly with signal.}
    \label{fig:synthetic}
\end{figure}

In order to illustrate how edge nodes and entropy information can
help in the visual analysis of spatio-temporal data, we have manufactured a 
data set by randomly placing $600$ points within a 2D unitary square, connecting 
the points using the Delaunay triangulation. The points correspond to the graph
nodes and the Delaunay links to the graph edges. All graph edges are
assumed to have unitary weight. A spatio-temporal signal
$f:V\times T\rightarrow\mathbb{R}$
is associated to the nodes of the graph. In each time slice the signal $f$ is equal to $1$ for nodes
lying inside a circle with radius 0.1 and centered in the position $(0.5+\delta_x,0.5+\delta_y)$,
otherwise, $f$ is set to $0$.
The pair $\delta_x$ and $\delta_y$ correspond to a random perturbation in the center of the circle, each ranging
in the interval $[-0.05,0.05]$.
The random perturbation prevents the signal to be exactly the same in each time slice.
A random noise in the interval $[-0.1,0.1]$ is also added to $f$ in each time slice.
One hundred time slices are generated using the procedure described above, however,
in $12$ time slices the center of the circle is further translated towards the top right or bottom left corner
of the unitary square domain, forcing an abrupt change in the signal in those time slices. 
Figure~\ref{fig:synthetic} shows the signal generated as described above.
Top left image corresponds to a time slice where the circle is centered close to $(0.5,0.5)$ while
bottom left and top right images correspond to time slices where the center of the circle has been 
shifted towards the top right or bottom left corner of the unitary square. For each time slice, 
the corresponding edge node configuration is illustrated in the inset on the bottom right.
 
\begin{figure*}[ht!]
\centering
    \includegraphics[width=0.98\linewidth]{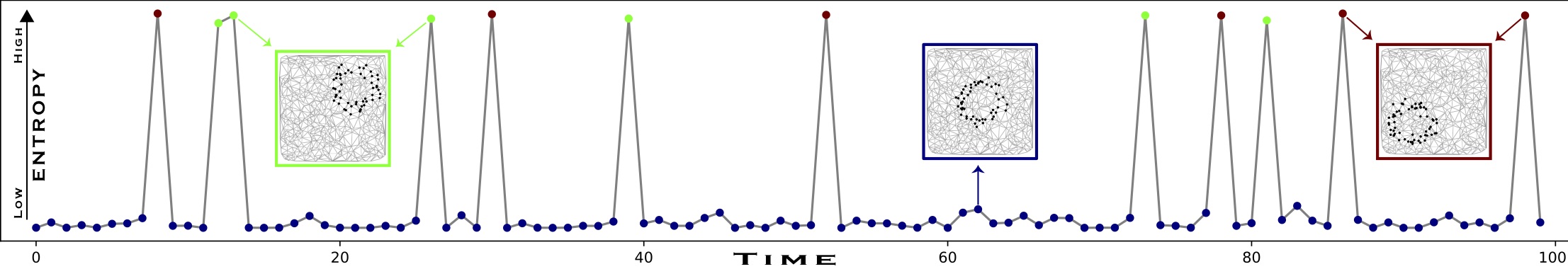}
    \caption{Entropy over time. The three patterns revealed correspond to $6$ time slices where
    the signal is shifted towards the top right corner, $6$ where the signal is shifted to the bottom left corner,
    and the remaining ones correspond to time slices where the signal is centered close to $(0.5,0.5)$.
    The $12$ time slices holding abrupt changes in the signal are clearly revealed by the entropy diagram.}
    \label{fig:synthetic_entro}
\end{figure*}

Figure~\ref{fig:synthetic_entro} presents the entropy diagram of the signal $f$ described above. 
Notice that the $12$ time slices
where the center of the circle is displaced towards the top and bottom corners of the unit square are 
revealed, as they have larger entropy values when compared to the remaining time slices.
Moreover, since the edge node configuration can be seen as a binary vector in an $n$-dimensional space,
we can cluster the high-dimensional vectors in order to identify time slices with similar edge node configuration. 
The color of the dots in Figure~\ref{fig:synthetic_entro} indicates the cluster each time slice
belongs to (we have clustered edge node configuration vectors in three groups using a simple k-means).
The red group corresponds to time slices where the center of the circles have been shifted towards the bottom left of the unit square, the green group are the time slices where the centers moved towards the top right corner.
Blue group gathers time slices where the center of the circles are close to $(0.5,0.5)$. 

Although simple, the synthetic example discussed in this section illustrates the potential 
of the GLoG filter and entropy diagram as visual analytic tools
for spatio-temporal data. Such usefulness will become more clear in the case 
studies presented in the following section.

%-------------------------------------------------------------------------
\section{Case Studies}
\label{sec:cs}
In this section we apply our methodology in two case studies involving real data. 
In both studies we strongly rely on the proposed GLoG filter and associated entropy diagram
to assist the visual analysis and identification of spatio-temporal patterns and phenomena. 
The first case study accounts for the analysis of taxi pick up during one week in downtown Manhattan - NYC. 
The second case study deals with crime data in the city of S\~ao Paulo - Brazil.
 
\begin{figure}[!t]
\centering
   \includegraphics[width=.975\linewidth]{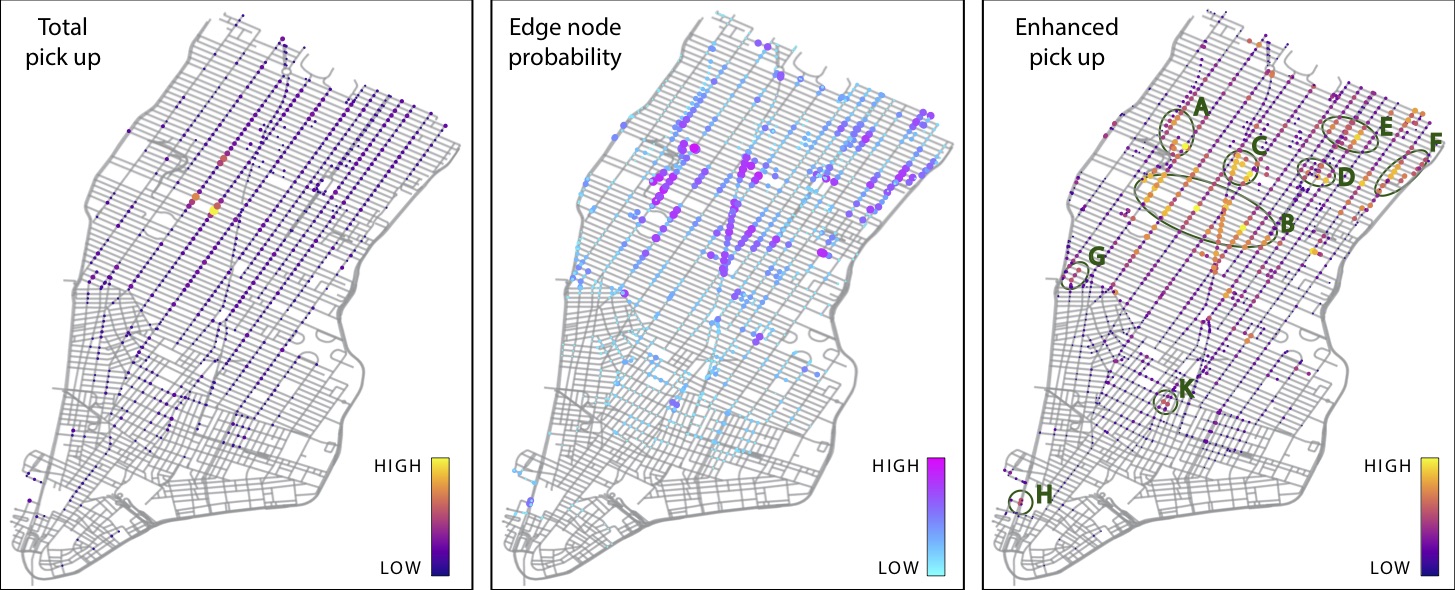}
    \caption{Total number of taxi pick ups (left), edge nodes probability $p_e$ (center), and enhanced total taxi pick up visualization (right).}
    \label{fig:total}
\end{figure}

\begin{figure*}[!t]
\centering
    \includegraphics[width=0.975\linewidth]{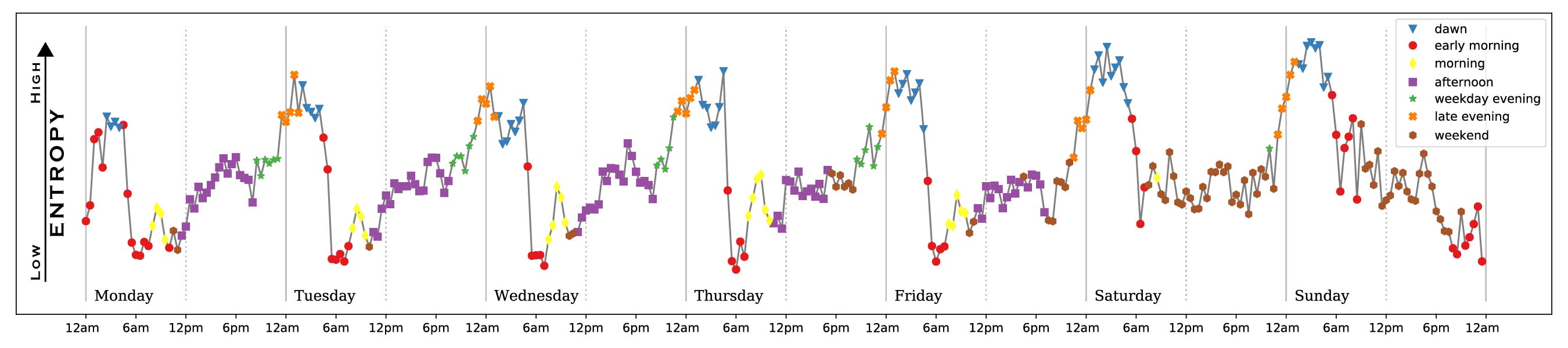}
    \caption{Taxi pick up entropy diagram. Entropy tends to increase along the day in weekdays, reaching its maximum around midnight.
    On Saturday, the gradual increase along the day is not observed, but an abrupt growth in the late evening is.
    On Sunday, entropy tends to decrease along the day.
    Therefore, taxi pick up is more random at dawn, specially in the weekends.
    Edge node configuration (see Figure~\ref{fig:nec}) of each time slice have been clustered in seven groups,
    revealing the different behavior of taxi pick up along the day.
    }
    \label{fig:taxi_entro}
\end{figure*}
\begin{figure*}[!t]
\centering
    \includegraphics[width=0.975\linewidth]{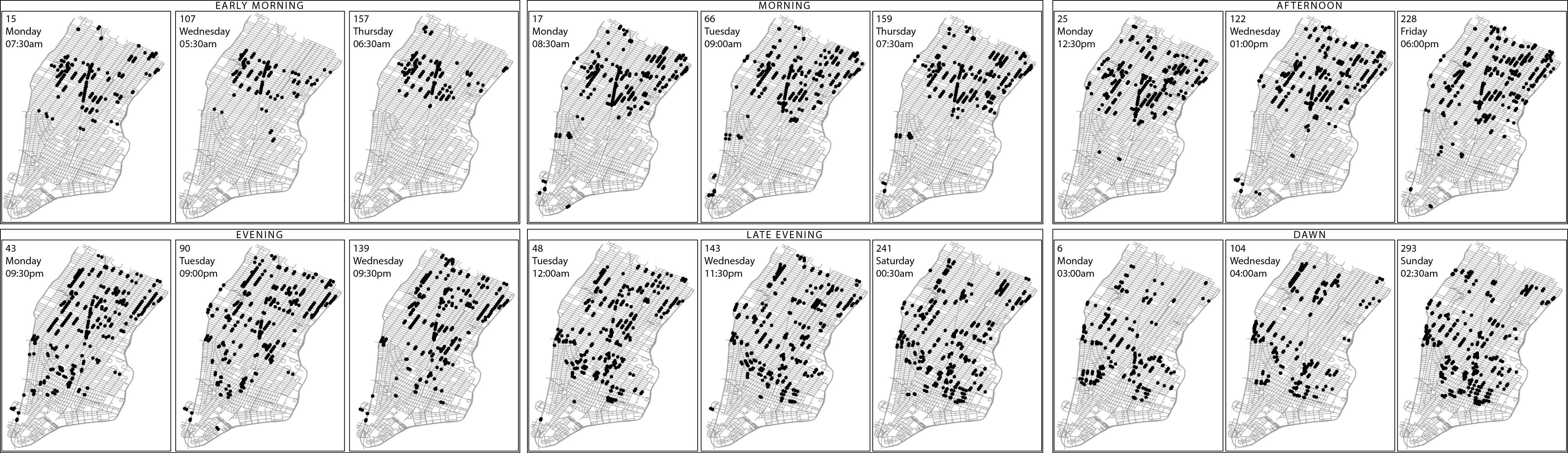}
    \caption{Example of edge node configuration along the day in weekdays. 
    The edge node configuration reveals the dynamics of the city,
    with a concentration of edge nodes in midtown in the early morning,
    extending to east side in the early afternoon, and moving to downtown in the evening.  
     }
    \vspace{-0mm}\label{fig:nec}
\end{figure*}

%\vspace{-0.3cm}
\subsection{NYC Taxi Data Analysis}
The graph domain in this case study is downtown Manhattan street network,
where street intersections correspond to graph nodes and the street blocks 
connecting the intersections make up the set of graph edges.
The graph of downtown Manhattan is made up of 4694 nodes 
and 6350 edges in each time slice. 
The Taxi data set~\cite{ferreira:tvcg:2013} contains information about one week of 
taxi pick up in each intersection of downtown Manhattan from August 11th to August 18th, 2013. 
The data was aggregated into half-hour intervals in each node, giving rise to 336 time slices.
The GLoG filter is applied to each time slice independently. 

Figure~\ref{fig:total} left shows the total number of taxi pick up in each node of the graph 
(the sum, over the 336 time slices, of the number of taxi pick ups in each node).
A few nodes are highlighted in the neighborhood of Penn Station,
a major train and metro station hub in NYC, and a few blocks uptown, close to Port Authority,
the main bus station terminal in Manhattan.
Since the number of taxi pick ups in those two regions are much larger than in other 
areas of the city, other important locations are overshadowed, becoming difficult to visualize 
regions where the number of taxi pick ups is relevant but not so large as 
in the neighborhood of Penn Station and Port Authority. 
Figure~\ref{fig:total} middle shows the probability $p_e(\tau_i)$ (see Eq.~\ref{eq:prob}) of each  
edge node $\tau_i$. 
Since the GLoG filter is scale invariant, it reveals
a number of nodes that are not highlighted in Figure~\ref{fig:total} left, all corresponding to 
nodes with high probability of being edge nodes.
Nodes presenting a not so high signal intensity but with high probability of being an edge node
are usually associated to interesting spatial events that deserve to be analyzed.
Figure~\ref{fig:total} right corroborates this claim, showing an %heat
enhanced total taxi pick up visualization generated by adding 
the probability $p_e$ to the normalized total number of taxi pick ups.
The combined visualization is the equivalent of what is called ``image enhancement'' in the 
context of image processing~\cite{shapiro2001computer}. 
As annotated in Figure~\ref{fig:total} right, besides Port Authority (A), the whole neighborhood
of Penn Station, including Korea Town (B), Times Square (C), Grand Central Station (D),
the luxurious hotel neighborhood (E), and the region that concentrates most of the consulates in NY (F)
are also visible in the enhanced visualization. 
Even smaller spots like the Meatpacking District (G) and the 9/11 memorial (H) show up after the enhancement process.
The region indicated by (K) is also interesting to analyze, as it corresponds to 
Jersey St., a very quiet two-block street in the middle of Soho where the number
of taxi pick ups is much smaller if compared against its neighborhood.
Notice that the enhanced visualization not only reveals important areas in Manhattan 
but also locations with almost no pick ups, which can indicate quiet spots or issues in
the traffic as bloched streets and accidents.
\begin{figure}[!t]
\centering
    \includegraphics[width=0.975\linewidth]{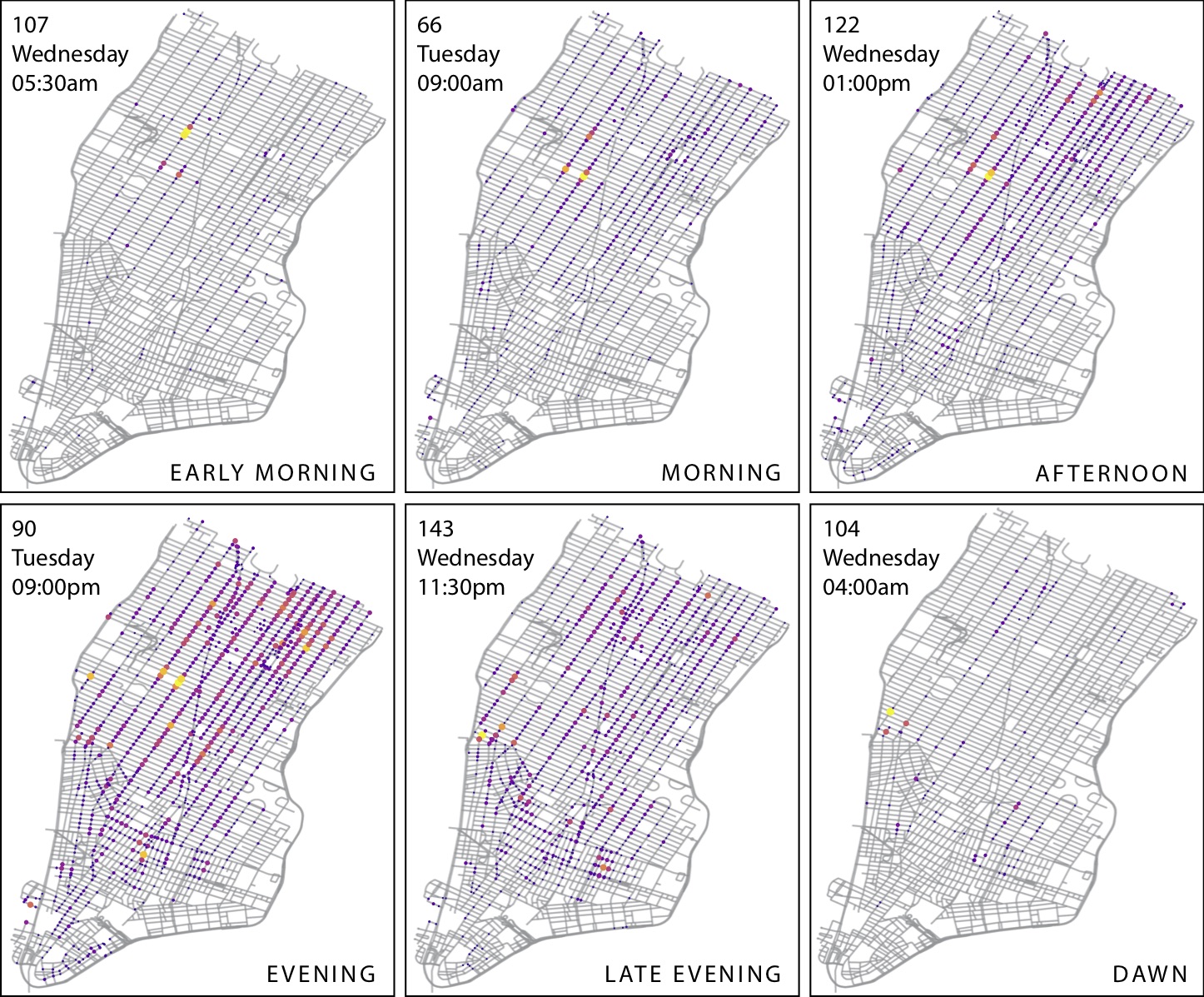} 
    \caption{Enhanced taxi pick up visualization
    for some of the time slices depicted in Figure~\ref{fig:nec}. 
    }
    \label{fig:s}
    \vspace{-0.5cm}
\end{figure}

{%\color{blue}
Figure~\ref{fig:taxi_entro} shows the taxi pick up \emph{entropy diagram}.
Notice that entropy behaves similarly in the weekdays,
reaching its maximum after midnight, dropping down at the dawn, presenting 
a local maximum right after 6am, increasing consistently along the day.
Weekend days also present maximum entropy at dawn,
however, the gradual increase along the day is not observed. In contrast, an abrupt growth 
is observed in late evening on Saturday while entropy decreases along the day on Sunday. 
Therefore, the entropy diagram shows that taxi pick up becomes more random at dawn, specially in the weekend. 
}

Edge node configuration associated to each time slice have been clustered in seven different
groups using $k$-means (as explained in Section~\ref{sec:synt}). 
The number of clusters has been defined by analyzing the silhouette coefficient~\cite{rousseeuw1987silhouettes}, 
which indicated the number $7$ as the ``ideal'' number of clusters to balance quality 
and number of groups. 
It is possible to see that the groups nicely split
the behavior of taxi pick up according to the periods of the day, 
indicating that the edge node configuration is capturing patterns of taxi pick up. 
Moreover, the particular behavior taking place during the weekend is also captured by the edge node configuration.
Figure~\ref{fig:nec} shows randomly chosen examples of edge node configurations from $6$ out of the $7$ groups. 
The \emph{Early Morning} group concentrates
edge nodes mostly in the neighborhood of main public transportation hubs such as Penn Station, Port Authority, 
Grand Central, and other smaller path train and metro station terminals.
The \emph{Morning} group shows that taxi pick up spread to northeast of downtown Manhattan, 
in the area of luxurious hotels and embassies. In the \emph{Afternoon} group, one can see that taxi pick up 
start to move downwards to 14th street, with a concentration around Union Square park. 
\emph{Weekday Evening} is characterized by a concentration of edge nodes in 
Soho and Meatpacking district, which are well known commercial areas with famous stores
markets and bars. In the \emph{Late Evening} group, edge nodes spread to East Village,
an area with many bars and restaurants that is known by its nightlife. Edge nodes in 
the \emph{Dawn} group are mostly concentrated in   
the south part of the island, presenting a reduction on the number of edge nodes in the northeast area.
As pointed out by the entropy diagram, late night and dawn have larger entropy, meaning that the
pattern of taxi pick up is more random during those periods.

The visualization of edge node configuration patterns depicted in Figure~\ref{fig:nec} 
reveals the dynamics of the city, where people tend to arrive in the public transportation terminals
during the morning, staying more concentrated in the north and northeast area during the day,
migrating to the south and east side of Manhattan in the evening and
late night to enjoy the bars and restaurants located on those areas.
Figure~\ref{fig:s} shows the enhanced visualization of 
taxi pick up in six time slices shown in Figure~\ref{fig:nec}, one for each group of edge node configuration.
Notice the agreement between the enhanced visualization of taxi pick up and the patterns of edge node configuration,
showing that the edge node configuration can be an interesting alternative when analyzing the behaviour of a signal
overtime, since even small variation on the signal can be identified as edge nodes. 

\subsection{Crime Data}
In this case study we apply the proposed methodology to analyze crime data in the city of S\~ao Paulo - Brazil. The data 
is available for downloading in the S\~ao Paulo Open Data repository (http://dados.prefeitura.sp.gov.br) 
and it provides, among others, information about date and time of each criminal event, the type of crime, 
and the census region where the crime event took place. In this study we
focused on eleven  years of data, from 2007 to 2017, accounting only for \emph{passerby robbery} as crime type.
Since the geolocation of each criminal activity is given by census regions, we have built a graph where nodes correspond to 
the census region and graph edges correspond to links connecting census regions that geographically intersect each other. 
Only census regions at a distance of less than five kilometers from the center of S\~ao Paulo are considered, 
resulting on a graph with $3,805$ nodes and $12,483$ edges.
We have applied a month aggregation to the data, resulting in $132$ time slices. 
\begin{figure}[!t]
\centering
    \includegraphics[width=0.975\linewidth]{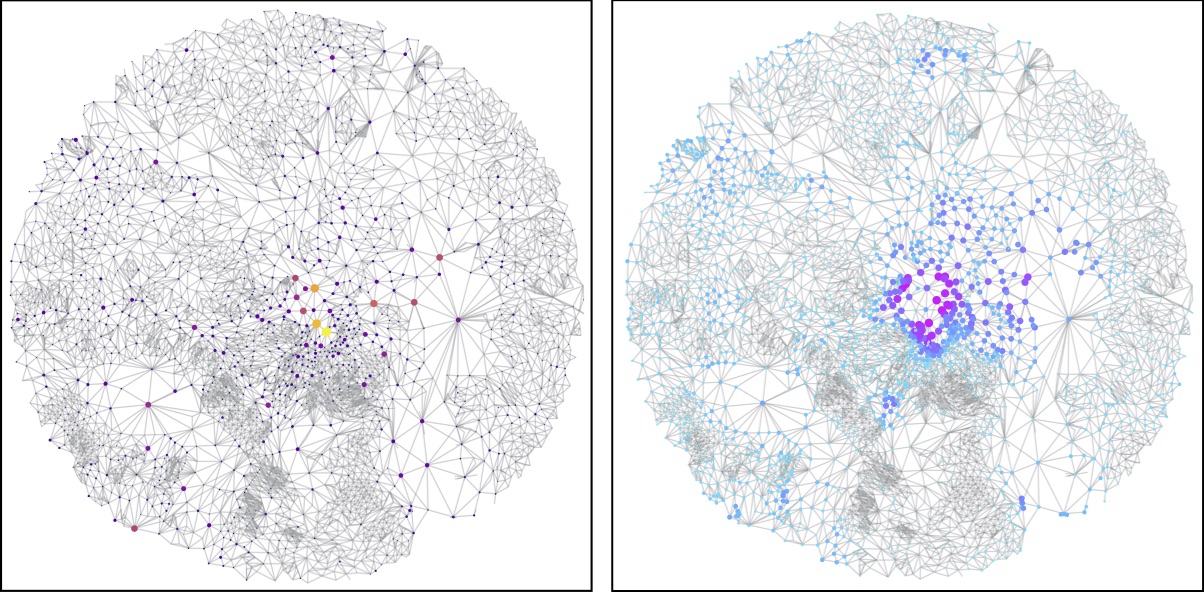}     
    \caption{Total number of passerby robbery from 2007 to 2017~(left). The probability $p_e(\tau_i)$
    of each node being an edge node~(right).}\vspace{-0.0cm}\label{fig:crime_total}
    \vspace{-0.5cm}
\end{figure}

\begin{figure*}[!t]
\centering
    \includegraphics[width=0.95\linewidth]{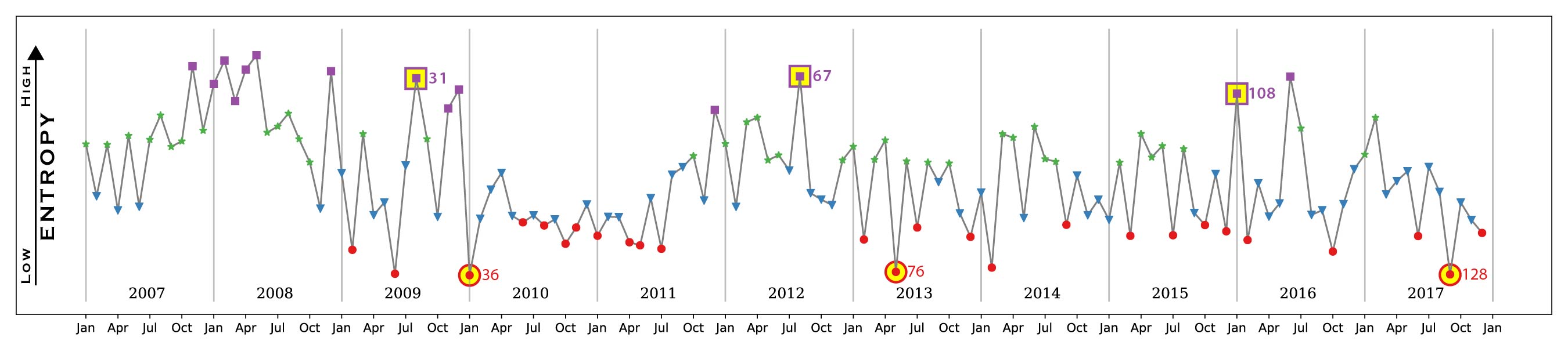}
    \caption{Entropy diagram of S\~ao Paulo passerby robbery. The highlighted picks correspond to some of the highest and lowest entropy 
    time slices.} \vspace{-0.2cm}
    \label{fig:crime_entro}
\end{figure*}

\begin{figure}[!t]
\centering
    \includegraphics[width=0.975\linewidth]{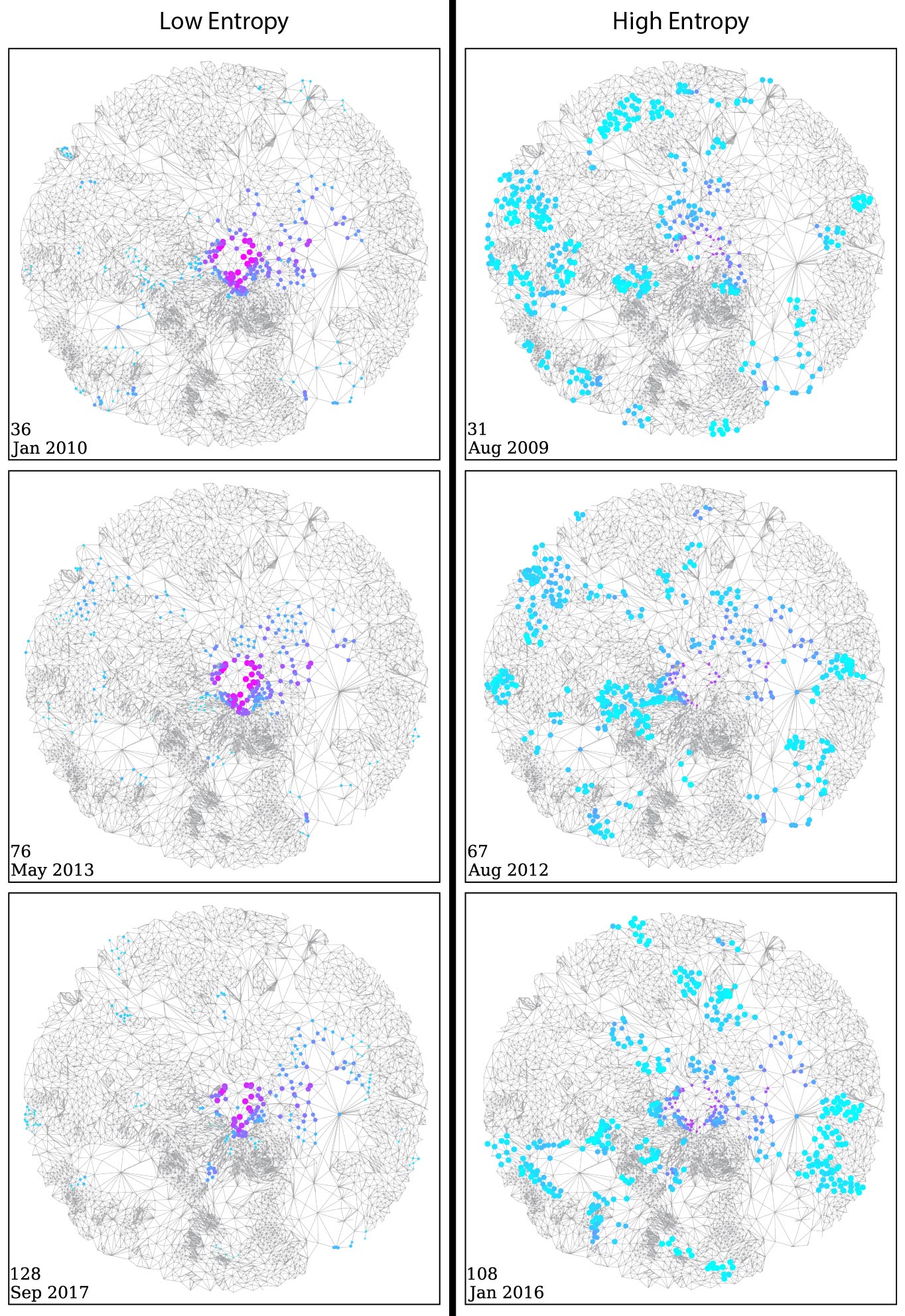}
    \caption{S\~ao Paulo crime: High probability edge nodes in low entropy type slices (left ) and low probability edge nodes 
    in high entropy time slices (right).
    Left column, purple points indicate \textbf{high probability} edge nodes, which are locations where criminal activity is common.
    Right column, blue points indicate \textbf{low probability} edge nodes, indicating neighborhoods where the criminal activity is unexpected.}
    \label{fig:crime_edges}
    \vspace{-0.5cm}
\end{figure}

Figure~\ref{fig:crime_total} shows a visualization of the total number of passerby robbery along the years (left) and the probability $p_e(\tau_i)$ (see Eq.~\ref{eq:prob}) of each node $\tau_i$ being
an edge node (right) (purple colors indicate larger probabilities). As in the case of NYC taxi pick up, the number of 
passerby robbery in the center of S\~ao Paulo is much larger than in other regions, overshadowing 
nodes and regions where the number of passerby robbery is not in the same magnitude as in the city center, but
still important to be identified and analyzed. 

The entropy diagram of S\~ao Paulo passerby robbery is depicted in Figure~\ref{fig:crime_entro}. 
Three high and three low entropy time slices are chosen to be investigated in detail (highlighted 
in the entropy diagram). High entropy time slices are associated to (edge node)
patterns that are less likely to happen.
In the context of crime data, this means that crimes are taking place in ``unexpected'' locations. 
In fact, an edge node configuration with high entropy tends to bear a considerable number of low probability ($p_e$) edge nodes. 
Therefore, by analyzing low probability edge nodes that show up in high entropy time slices one can identify 
locations where the number of passerby robbery changes abruptly, which can indicate the action of gangs 
or emerging of juvenile offenders in those location or even a sudden reduction on crime rates, which is also
interesting to analyze.
The images on the right column in Figure~\ref{fig:crime_edges} shows low probability (probability lower the $0.5$) 
edge nodes present in the high entropy time slices highlighted in Figure~\ref{fig:crime_entro}. 
Notice that a number of locations and neighborhoods far from the city center are evident.
By analyzing the number of crime events in those locations we noticed that they do not present a 
large number of passerby robbery but, in the pointed high entropy time slices, 
the number of criminal events increases substantially. 
In contrast, the edge node configuration associated to low entropy time slices tend to reveal (edge node) 
patterns that are probable to happen, thus indicating neighborhoods where criminal activities are expected.
Left column in Figure~\ref{fig:crime_edges} shows high probability edge nodes present 
in the low entropy time slices indicated in Figure~\ref{fig:crime_entro}. 
It is easy to see that the most probable edge nodes are
located in downtown S\~ao Paulo, indicating a frequency of passerby robbery in that area. 
The analysis above shows the potential of GLoG as a basic tool to support visual analytic tasks,
mainly in applications where scale invariance is a basic requirement, as is the case of crime analysis.  

%-------------------------------------------------------------------------
\section{Discussion and Limitations}

The case studies presented in Section~\ref{sec:cs} shows the effectiveness
of the proposed GLoG filter and entropy diagram 
to support the visual analysis of spatio-temporal data.
The entropy diagram reveals the degree of randomness of each time slice, allowing an
intuitive visual identification of predictable and unpredictable time slices
in terms of their edge node configuration.

The edge node configuration is, by itself, a human interpretable signature from which
one can understand the spatial behavior of a signal. With the help of clustering schemes, 
edge node configuration can uncover time intervals where a signal presents similar behavior,
thus helping users to understand spatial phenomena over time.

{Our methodology demands essentially two parameters, $\sigma$, which controls the 
strength of the Gaussian filter, and the threshold used to filter out weak edge node pairs.
The latter is defined in terms of the standard deviation of the distribution of edge node pair scores.
In the case studies we fixed $\sigma=3$ and the edge node pair threshold in the third quartile,
as those values resulted on interesting analysis and visual patterns.
However, those parameters impact in the edge node configuration and thus in the entropy 
and cluster analysis. 
The larger the $\sigma$ and the threshold smaller is the number of edge nodes tend to be present in the edge node configuration.
Although we have not experienced any difficulties to set those parameters
in the two case studies discussed in Section~\ref{sec:cs}, 
that might not always be the case, and a fine tuning can be necessary 
depending on the application.}

Regarding computational times, the whole process, including 
the computation of the largest eigenvalue (involved in the Chebyshev approximation),
GLoG filtering, edge node pairs identification and threshold, takes only a few seconds,
which is pretty reasonable.
For instance, the whole process took less than $5$ seconds in both case studies.

The proposed methodology opens a wide set of future directions. For instance,
it would be interesting to investigate techniques able to track edge nodes over time 
in order to visualize how the spatial ``boundaries'' evolves, allowing a better understanding of the 
dynamics of the data. Another interesting problem is how to
build upon boundary detection to design spatial segmentation techniques that respect
the edge nodes. This kind of segmentation has been successfully employed in image processing and
computer vision. Properly segmenting spatial locations based on data is 
an major problem in visualization that can greatly benefit from the methodology proposed
in this work. 
The  GLoG filter can also be combined with other feature extraction 
mechanisms such as topological methods, allowing the analysis of ''extremal'' and
transitional data locations. 
Moreover, the proposed methodology can also be applied to problems 
beyond geo-spatial data, as for example high-dimensional data analysis, mainly through
dimensionality reduction.

%-------------------------------------------------------------------------
\section{Conclusion}
In this work we have introduced a novel methodology to process
and visualize spatio-temporal data based on the concept of boundary detection and entropy diagram. 
The proposed methodology strongly relies on spectral filtering from graph signal processing theory,
which enables the precise definition of a Laplacian of Gaussian boundary detection filter that operates in graph domains.
Edge nodes computed by the proposed machinery can be used as feature vectors as well as
to define the notion of entropy of time slices, allowing the identification of low and high 
probable time slices. Visual analytic tasks involving synthetic and real data sets 
show the effectiveness and usefulness of the propose methodology to support visualization applications. Moreover, the proposed methodology can also be combined with other feature extraction methods so as to
capture different facts of time-varying data, opening a number of possibilities for
future work. 

\ifCLASSOPTIONcompsoc
  \section*{Acknowledgments}
\else
  \section*{Acknowledgment}
\fi

This work was supported by: the Moore-Sloan Data Science Environment at NYU;
NASA; NSF awards CNS-1229185, CCF-1533564, CNS-1544753, CNS-1626098, CNS-1730396, CNS-1828576;
302643/2013-3 CNPq-Brazil and 2016/04391-2 and 2014/12236-1 S{\~a}o Paulo Research Foundation (FAPESP) - Brazil.
C. T. Silva is partially supported by the DARPA MEMEX and D3M programs.
Any opinions, findings, and conclusions or recommendations expressed in this material are those of the authors
and do not necessarily reflect the views of DARPA and S{\~a}o Paulo Research Foundation.

\ifCLASSOPTIONcaptionsoff
  \newpage
\fi



%\bibliographystyle{IEEEtran}
%\begin{thebibliography}{1}
%\end{thebibliography}
\bibliographystyle{IEEEtran}
\bibliography{refs}


%\begin{IEEEbiography}[{\includegraphics[width=1in,height=1.25in,clip,keepaspectratio]{gustavo.jpg}}]
%{Luis Gustavo Nonato}
% ....
%\end{IEEEbiography}

%\begin{IEEEbiography}[{\includegraphics[width=1in,height=1.25in,clip,keepaspectratio]{bio/gustavo}}]{Luis Gustavo Nonato}
\begin{IEEEbiography}{Luis Gustavo Nonato}
...
%received the PhD degree in applied mathematics from the
%Pontif\'{\i}cia Universidade Cat\'{o}lica do Rio de Janeiro, Rio de Janeiro ---
%Brazil, in 1998. He is a full professor at the Instituto de Ci\^{e}ncias
%Matem\'{a}ticas e de Computa\c{c}\~{a}o (ICMC) --- Universidade de S\~{a}o Paulo (USP) ---
%Brazil. He spent a sabbatical leave in the Scientific Computing and Imaging
%Institute at the University of Utah from 2008 to 2010. Besides having
%served in several program
%committees, including IEEE Visualization and Eurovis, he was a member of
%the editorial board of Computer Graphics Forum and the president of the
%Special Committee on Computer Graphics and Image Processing of Brazilian
%Computer Society. Currently Dr.\ Nonato leads the Visual and Geometry
%Processing Group at ICMC-USP\@. 
\end{IEEEbiography}

%\begin{IEEEbiographynophoto}{Fabiano Petronetto do Carmo}
% ...

%\newpage
\begin{IEEEbiography}{Fabiano Petronetto do Carmo}
...
%received his PhD degree in Applied Mathematics 
%from Pontif\'{\i}cia Universidade Cat\'{o}lica do Rio de Janeiro (PUC-Rio) in 2008.
%After a short time working in research projects supported by Petrobras
%(Brazilian Oil Company), he started as professor of the Mathematics
%Department at Universidade Federal do Esp\'{\i}rito Santo in 2009. 
%His research interests are in development of meshless methods using 
%the particle method Smoothed Particle Hydrodynamics (SPH).
%He has collaborated with the Visual and Geometry Processing Group at
%Instituto de Ci\^{e}ncias Matem\'{a}ticas e de Computa\c{c}\~{a}o (ICMC-USP), where
%he currently holds a post-doc (2013--2015).
\end{IEEEbiography}

\begin{IEEEbiography}{Claudio Silva}
...
\end{IEEEbiography}


%\vfill


\end{document}